\documentclass[10pt,twocolumn]{IEEEtran}
\usepackage{fancyhdr}
\usepackage{graphicx}
\usepackage{amsfonts}
\usepackage{times}
\usepackage{latexsym}
\usepackage{amssymb}
\usepackage{amsmath}
\usepackage{cite}
\usepackage{verbatim}
\usepackage{subfigure}
\usepackage{multirow}
\usepackage{lastpage}
\usepackage{calc}
\usepackage{eso-pic}
\usepackage{tabularx}
\usepackage{ifthen}

\def\bb0{{\mathbb{0}}}

\def\bb{{\mathbf{b}}}

\def\b0{{\mathbf{0}}}

\def\sf0{{\mathsf{0}}}

\def\SINR{\mathrm{SINR}}

\usepackage{epstopdf}
\usepackage{graphicx,subfigure}
\usepackage{enumerate}
\usepackage[binary-units,detect-weight=true, detect-family=true]{siunitx}
\DeclareSIUnit \bitspersecond {bps}

\definecolor{deeppink}{rgb}{1.0, 0.08, 0.58}
\definecolor{green(ncs)}{rgb}{0.0, 0.62, 0.42}

\begin{document}
\title{Non-Stationarities in \\ Extra-Large Scale Massive MIMO}

\author{\normalsize
Elisabeth De Carvalho,~\IEEEmembership{Senior Member,~IEEE,}
Anum Ali,~\IEEEmembership{Student Member,~IEEE,}
Abolfazl Amiri,~\IEEEmembership{Student Member,~IEEE,}
Marko Angjelichinoski,~\IEEEmembership{Member,~IEEE,}
Robert W. Heath Jr.,~\IEEEmembership{Fellow,~IEEE}
\thanks{
E. de Carvalho and A. Amiri  are with the Department of Electronic Systems, Aalborg University, 9220 Aalborg, Denmark
(e-mail: \{edc,aba\}@es.aau.dk).
Marko Angjelichinoski is with Department of Electrical and Computer Engineering, Duke University, Durham, NC 27708, USA
(e-mail: marko.angjelichinoski@duke.edu).
A. Ali and R. W. Heath Jr. are with the Wireless Networking and Communications Group, Department of Electrical and Computer Engineering, The University of Texas at Austin, Austin, TX 78712, USA (e-mail:\{anumali,rheath\}@utexas.edu).}
\thanks{This material is based upon work supported in part by TACTILENet (Grant no. 690893), within the Horizon 2020 Program, by the Danish Council for Independent
Research (Det Frie Forskningsr\aa d) DFF-701700271, and the National Science Foundation under Grant No. ECCS-1711702.
}
}

\maketitle
\begin{abstract}
Massive MIMO, a key technology for increasing area spectral efficiency in cellular systems, was developed assuming moderately sized apertures. In this paper, we argue that massive MIMO systems behave differently in large-scale regimes due to spatial non-stationarity. In the large-scale regime, with arrays of around fifty wavelengths, the terminals see the whole array but non-stationarities occur because different regions of the array see different propagation paths. At even larger dimensions, which we call the extra-large scale regime, terminals see a portion of the array and inside the first type of non-stationarities might occur. We show that the non-stationarity properties of the massive MIMO channel change several important MIMO design aspects. In simulations, we demonstrate how non-stationarity is a curse when neglected but a blessing when embraced in terms of computational load and multi-user transceiver design.
\end{abstract}

\section{Introduction} \label{sec:intro}
 
\begin{figure*} [!ht]
\centerline{
\includegraphics[width=0.8\linewidth]{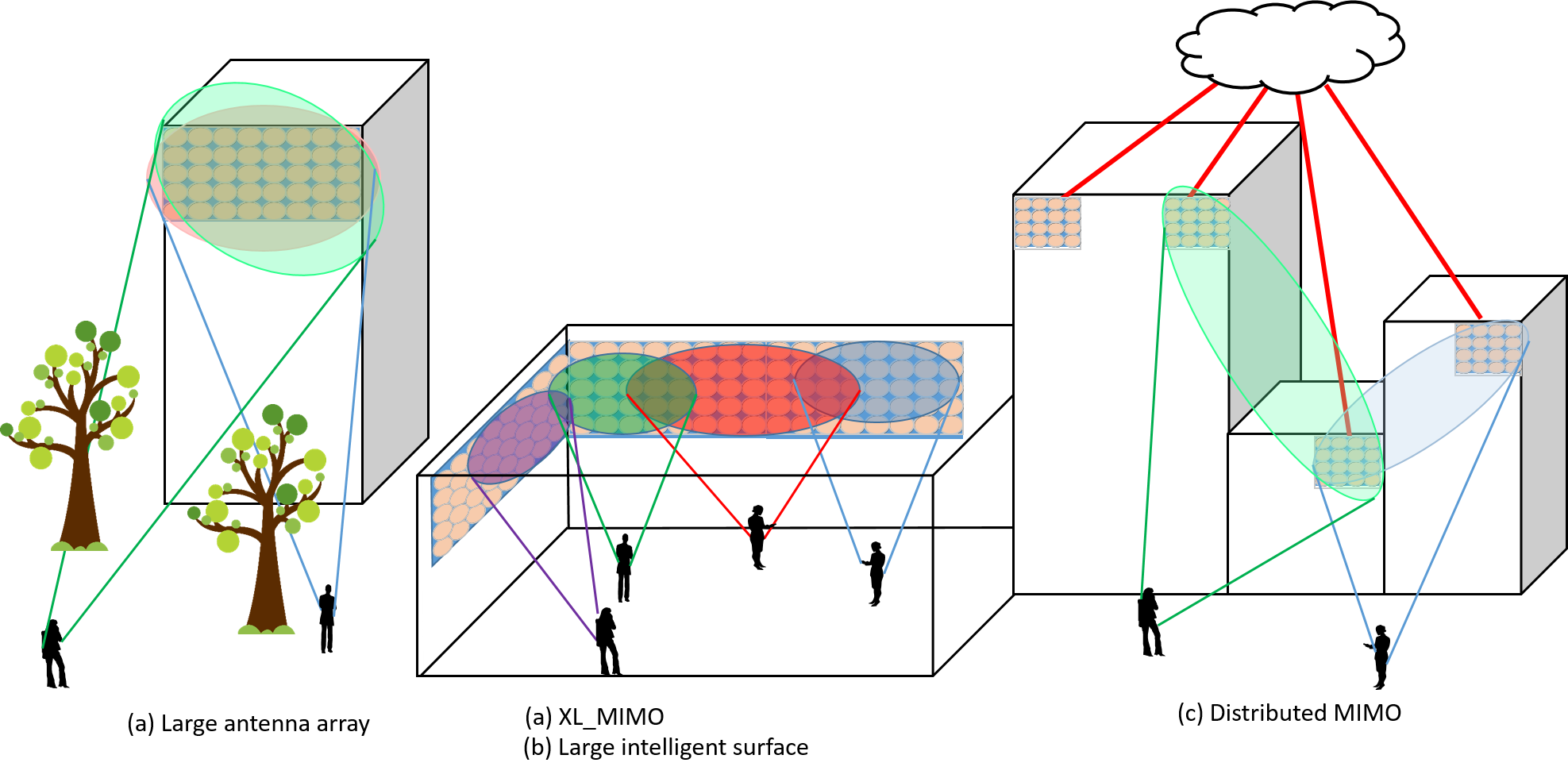}
   }
\caption{The ways to create larger apertures in a massive MIMO. (a) Antenna array with large dimension. (b)  Antenna array with extra-large dimension. (c) Large intelligent surface. (d) Distributed antenna system. }
\label{fig:city}
\end{figure*}
Massive multiple-input-multiple-output (MIMO) is a key technology in 5G wireless communication systems in sub-6 GHz bands. It is characterized by the use of many antennas at the base station serving many terminals simultaneously. In current cellular deployments, massive MIMO will likely be implementing compact planar arrays. The small footprint leads to reduced infrastructure costs. Even with a large number of antennas, though, compact design does not expose enough spatial dimensions. 

{Spatial dimensions are essential in uncovering the fundamental properties of massive MIMO: channel hardening, asymptotic inter-terminal channel orthogonality, and large array gains. 
Increasing the array dimension contributes to achieving the performance gains originally promised by massive MIMO and providing high data rates when the number of terminals is much smaller than the number of antennas. Increasing the array dimension further allows the support of high data rates to a much larger number of terminals. 
Distributing the arrays across a building, for example, allows for cost-efficient implementation of an extremely large array while bringing other benefits such as better coverage.}

The impact of the array dimension has motivated new types of deployment where the dimension of the arrays is pushed to the extreme. 
Such arrays would be integrated into large structures, for example
along the walls of buildings in a mega-city, in airports,
large shopping malls or along the structure of a stadium~\cite{Medbo:5G-channel-modelling:14,MarCarNie:Measurements-XL-MIMO:14} (see Fig.~\ref{fig:city}) and serve a large number of devices. The dimension of the arrays could attain several tens of meters. This type of deployment is considered an extension of massive MIMO with an implementation based on discrete antenna elements. 
We refer to this extreme case as extra-large scale massive MIMO (XL-MIMO). We argue in this paper that XL-MIMO should be considered a distinct operating regime of massive MIMO with its unique challenges and opportunities. 

When the antenna arrays reach such a large dimension, spatial non-wide sense stationary properties appear along the array. Different parts of the array may have different views of the propagation environment, observing the same channel paths with different power, or different channel paths~\cite{GaoTufEdf:MaMIMOModels:13}.
When the dimension of the array becomes extremely large, different parts of the array may also view different terminals as the energy of each terminal is focused on a portion of the array, called visibility region (VR). 
As the array dimension increases, the performance for each terminal is limited by its VR, i.e., the effective array dimension viewed from the array. However, the ability to serve multiple terminals with high data rates is highly enhanced, hence bringing benefits in crowded scenarios. 

Wireless communications involving large electromagnetic elements is an emerging concept. The term  Large Intelligent Surface (LIS) has appeared recently and denotes generically a large electromagnetic surface~\cite{HuRusEdf:LIS:2018} that is active and hence possesses communication capabilities. 
Another possible implementation of very large arrays is through radio stripes as described in~\cite{Larsson:Ubi:2018} that can be easily attached to existing construction structures and are connected to a central unit to form a distributed cell-free system.
Interestingly, research is also focusing on passive large electromagnetic surfaces~\cite{LiaNiePit:MetaSurf:2018}. A passive LIS acts as a reflecting surface that changes the properties of the incoming electromagnetic waves. It acts as a relay to enhance the propagation features of the reflected waves. 

In this article, we focus on discrete arrays of antennas, 
not continuous surfaces, and the effect of non-stationary properties along the array. 
Our emphasis is on VRs and their impact on performance and transceiver design. 
The primary differentiating feature from stationary massive MIMO is that the terminals have overlapping  VRs with an inter-terminal interference pattern that changes along the array. 
Non-stationarity is accounted for in the performance assessment of linear multi-terminal transceivers and design of hybrid analog-digital beamforming and serves as the main tool to alleviate the transceiver computational complexity.


\section{Types of spatial non-stationary regimes}

Fig.~\ref{fig:city} gives an overview of the types of deployments considered in this paper and the ways to create larger apertures for XL-MIMO. 

\begin{enumerate}[(a)]

\item An antenna array of large or extra-large dimension: typically embedded in a building of large dimension
\cite{Medbo:5G-channel-modelling:14}. 

\item  Large intelligent surface: a generic term for a large electromagnetic surface \cite{Medbo:5G-channel-modelling:14}. 
A possible implementation is with a discrete array of antennas (as in case (a)) but possibly other material.  

\item  Distributed antenna system:
cooperating antennas or arrays of antenna units placed at distant geographical locations~\cite{TruHeath:asilomar:2013}. 
\end{enumerate}

\begin{figure}
\centering     
\subfigure[Stationary massive MIMO ]{\label{fig:a}\includegraphics[width=0.9\columnwidth]{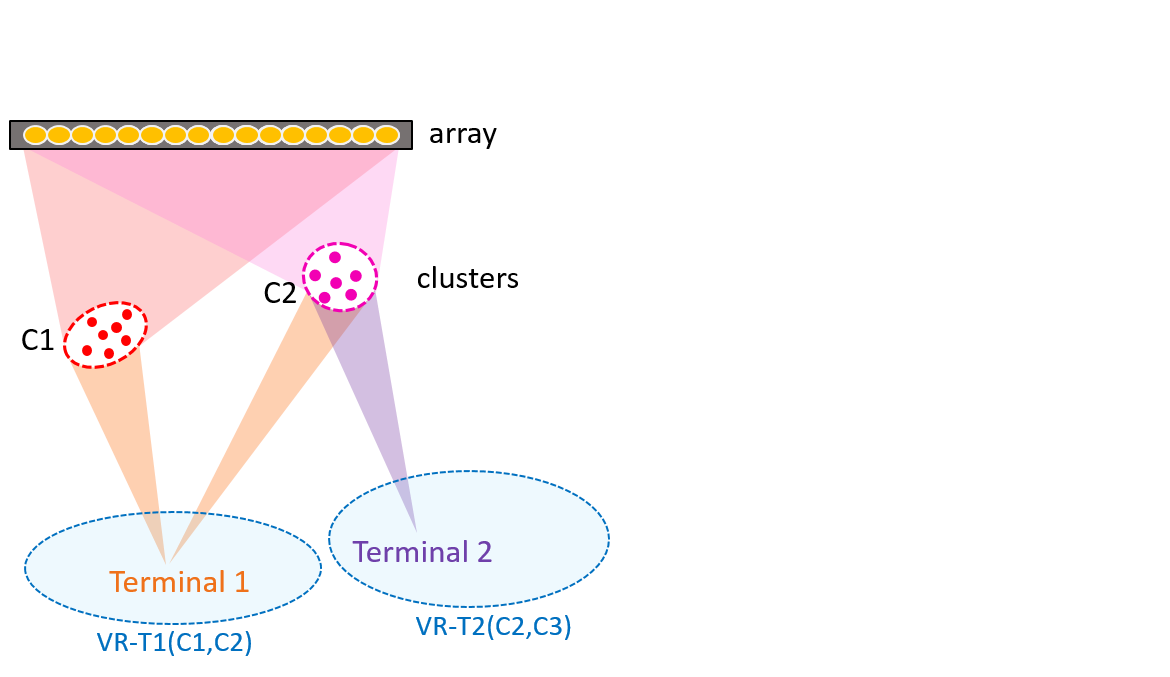}}
\subfigure[Large scale MIMO: clusters are visible from a portion of the array]{\label{fig:b}\includegraphics[width=0.9\columnwidth]{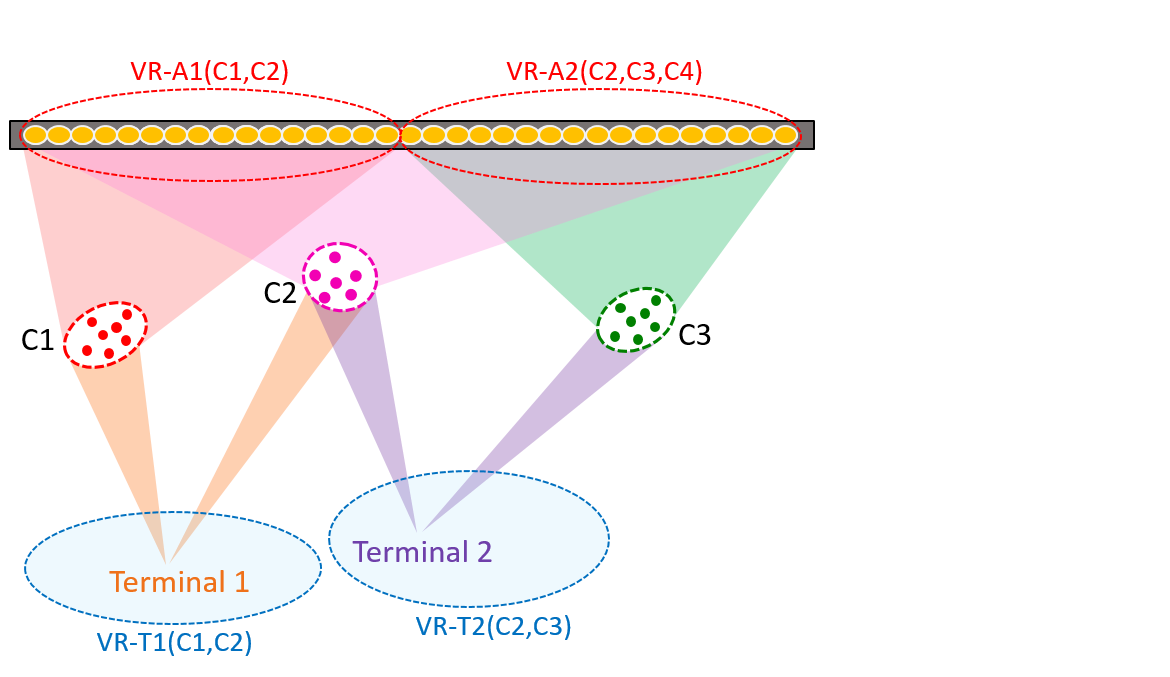}}
\subfigure[Extra-large scale MIMO: terminals are visible from a portion of the array]{\label{fig:c}\includegraphics[width=0.9\columnwidth]{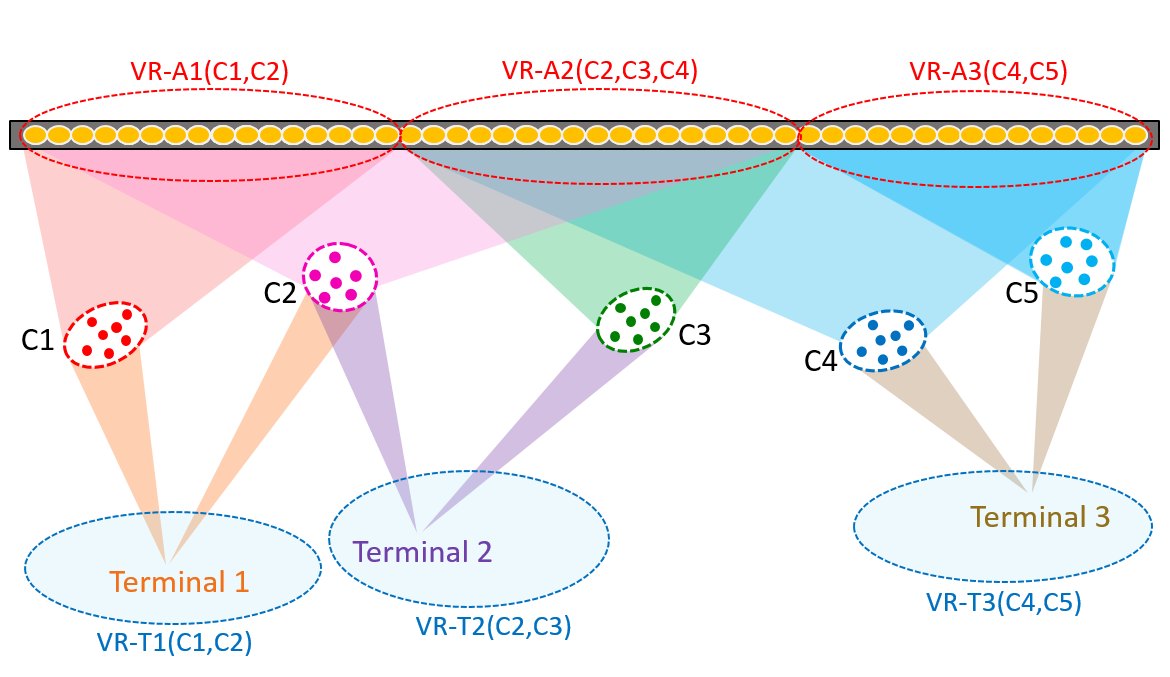}}
\caption{Three MIMO scales}
\label{fig:models}
\end{figure}

To illustrate the spatial non-stationarity properties in massive MIMO, we rely on a cluster-based channel model. 
Fig.~\ref{fig:models} depicts a conventional massive MIMO channel model that is spatially stationary, along with two types of spatial non-stationarities defined according to the concept of VR along an antenna array. 
The concept of VR was introduced in the COST 2100 channel model~\cite{LiuOestPou:COST2100:2012}. 
{In its original definition, a VR is a terminal geographical area. When the terminal is located in this area, it sees a given set of clusters. This is the set of clusters associated with the VR. When it moves out of the VR, the terminal sees a different set of clusters. 
We extend the concept of VR to denote a portion of the array from which a given set of clusters is visible. We distinguish between VRs in the terminal domain VR-T and in the array domain VR-A. }

\subsection{Large-scale massive MIMO} \label{sec:large}

{
The L-MIMO regime applies when different sets of clusters are visible from different portions of the array and the whole array is visible by all terminals. In general, this implies that the terminals are at a significant distance from the array. 
Fig.~\ref{fig:models}(b) illustrates a simple case where the array is divided into two disjoint VR-As.
}
This regime was highlighted in an early measurement~\cite{GaoTufEdf:MaMIMOModels:13} involving a long array of 7.4 meters in a courtyard where, at a different portion of the array, different propagation paths were measured. 
\subsection{Extra-large scale massive MIMO} \label{sec:extraLarge}

{The XL-MIMO regime applies when different sets of clusters as well as different sets of terminals are visible from different portions of the array. 
The main difference with the L-MIMO regime is that the terminals are much closer to the array (or the array is much larger). 
As seen in Fig.~\ref{fig:models} (c), one can define another type of VR: the portion of the array that is visible from a given terminal. For example, the VRs of terminal~1 along the array includes VR-A1 and VR-A2.  
}

\begin{figure}
\centering     
\subfigure[Average received power (dBm) in a 64 antenna array made of 8-antenna units (y-axis). The received power is averaged over the small movements of the terminals.]{\label{figRxPower:a}\includegraphics[width=0.9\columnwidth]{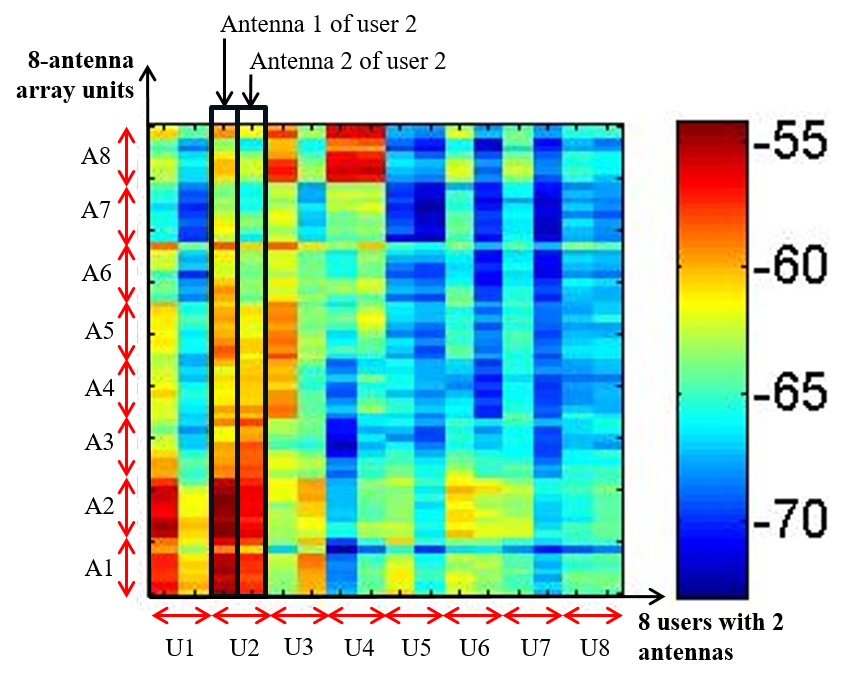}}
\subfigure[The array is 6 meters long. Eight terminals (y-axis) holding a  2-antenna device are around 2 and 6 meters from the array and move in a square of 1 square meter.]{\label{figRxPower:b}\includegraphics[width=0.9\columnwidth]{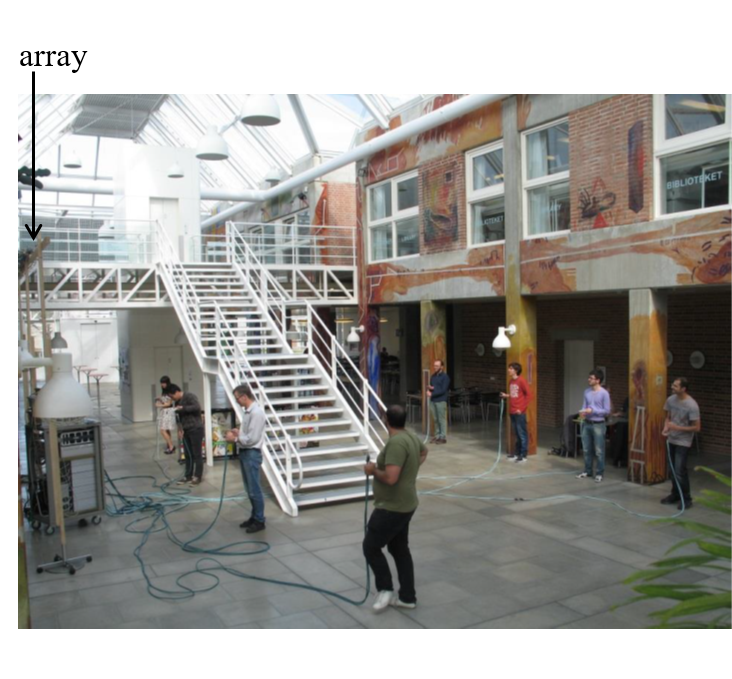}}
\caption{The measurement set-up and results for XL-MIMO~\cite{MarCarNie:Measurements-XL-MIMO:14}.}
\label{fig:RxPower}
\end{figure}

Aalborg University initiated a measurement campaign specifically dedicated to XL-MIMO~\cite{MarCarNie:Measurements-XL-MIMO:14} in a large indoor venue. 
Fig.~\ref{figRxPower:a} shows a striking result from the campaign that illustrates the complexity of the propagation environment. 
The massive array is six-meter long and comprises 64 antennas. It is placed along a wall on a line parallel to the floor and made of units of eight antennas. 
Eight terminals, around three meters apart, holding a two-antenna device are located at 2 and 6 meters in front of the array and send uplink signals as shown in Fig.~\ref{figRxPower:b}. 
Fig.~\ref{figRxPower:a} displays the average receive power of the channel when the terminals move locally. 
First, we observe very large variations of the power across the array, more than 10dB and different patterns for the two signals coming from the same device. 
Terminals 5 to 8 are located behind a stair case, which brings an attenuation of the signal visible from a portion of the array. 

\subsection{Distributed massive MIMO} \label{sec:distributed}
XL-MIMO can be seen as a special case of distributed massive MIMO where the whole set of arrays is collocated. Especially in a dense distribution, the same kind of model holds where clusters, as well as terminals, are visible from a subset of the arrays.
\subsection{Impact on key channel assumptions} \label{sec:assumption}

The non-stationary properties of arrays of very large dimension impose a departure from the conventional channel models, especially the widely used correlated channel model.  This model assumes that the channel has a centered Gaussian distribution with a covariance matrix that reflects stationary properties in the correlation among antennas as well as the propagation. While the Gaussian assumption might still hold, the most basic modification on the channel assumptions is that the average channel gain varies along the array. 
A cluster-based geometric channel model reflects more appropriately the source of non-stationarity, i.e. cluster VRs. The major change compared to traditional models is in the expression of the steering vectors.
First, near the array, the phase of each element should account for a spherical wave modeling as the planar wave approximation is not valid anymore. Second, the amplitude of each element varies. This is due to the path loss along the array as well as the interplay between clusters and obstacles in the environment as
different portions of the spherical wavefront might experience different propagation characteristics. 
The main drawback of this modeling is that it depends on the position of the clusters and terminals relative to the array, which makes it scenario-dependent and increases its complexity.

A simplification consists in decomposing the array in sub-arrays in which the channel is approximated as stationary. This model can be enhanced by adding a transition zone between the sub-arrays~\cite{martinez2016geometry}. This type of assumption can facilitate performance analysis of XL-MIMO systems~\cite{Ali2018Linear}. 
It motivates multi-antenna processing based on sub-arrays where the sub-array processing is adapted to the non-stationarity patterns.  
\section{Exploiting spatial non-stationarity} \label{sec:exploiting}
This section advocates that non-stationary properties, with a focus on VRs, should be accounted in performance assessment as well as transceiver design. We provide performance bounds of the zero-forcing (ZF) precoder using a simple VR model. Further, we demonstrate that VRs should be taken into account when designing hybrid analog-digital precoders/combiners. Finally, we exploit VRs, i.e., array regions where signals have low power, to design low complexity receivers. 

\subsection{Performance bounds} \label{sec:performance}

{Only a handful of studies have been conducted to study the impact of non-stationarity on the performance of massive MIMO systems. Channel capacity is studied in~\cite{Zhou2015Spherical} for a spherical wave-front based LOS channel model, while cluster non-stationarity visibility along the array is treated in~\cite{Li2015Capacity}. }
The focus of this section is on the impact of array VRs associated with the terminals and how it compares to the conventional stationary case. 
In~\cite{Ali2018Linear}, a simple non-stationary massive MIMO channel model was proposed so that it is conducive to the analysis of the effect of VRs. This model is employed to assess the performance of simple linear multi-terminal precoders (conjugate-beamforming (CB) and ZF precoders).

The channel of a given terminal is modeled as stationary within its VRs and is set to zero outside of the VR. Though the model was developed for correlated channels (see~\cite{Ali2018Linear}), here, we limit our discussions to independent and identically distributed channels, for simplicity, and ZF precoding. 
Consider a MIMO broadcast channel with $K$ single-antenna terminals served by a BS with $M$ antennas.
The $\SINR$ of terminal $k$ for ZF precoding averaged over stationary channels has a well-known expression. However,  VR-based channels are not easily amenable to analysis. 
It is possible however to find an approximation of the SINR, valid in asymptotic conditions, as a function of the VR size of each terminal and the size of the overlap regions. For simplicity, we assume that the terminals have the same VR size equal to $D$ antennas and total transmit energy per VR is equal to $M$. 
{We examine the worst and best case terminal configuration. The SINR can be written as 
$\frac{\rho}{K}(M-L(K,M,D))$ where the loss term $L(K,M,D)$ differs in each case. The term $\rho$ is the transmit signal-to-noise ratio.}

In the worst case, all the terminals have completely overlapping VRs - i.e., they receive the signal from the same $D$ antennas - the inter-terminal interference is high. 
In the best-case, inter-terminal interference is minimized asymptotically for all  $K$ terminals. 
The terminals are grouped in  $M/D$ groups where each group contains  $\frac{KD}{M}$ terminals. Hence,
 there are $\frac{KD}{M}-1$ interfering terminals for any terminal $k$ with an overlapping zone of $D$ antennas. 

The SINRs all scale as $M/K$ and differ in lower order quantities. 
The best-case non-stationary scenario results in better performance than the stationary case. It reaches its largest value when  $\frac{M}{D}$ is large, i.e., for small VRs or non-overlapping VRs. 
The worst-case non-stationary scenario results in worst performance than the stationary case. The smaller the VR of the terminal, the more SINR~loss compared to the stationary case. 

\begin{figure}[h!]
\centering
        \includegraphics[width=0.45\textwidth]{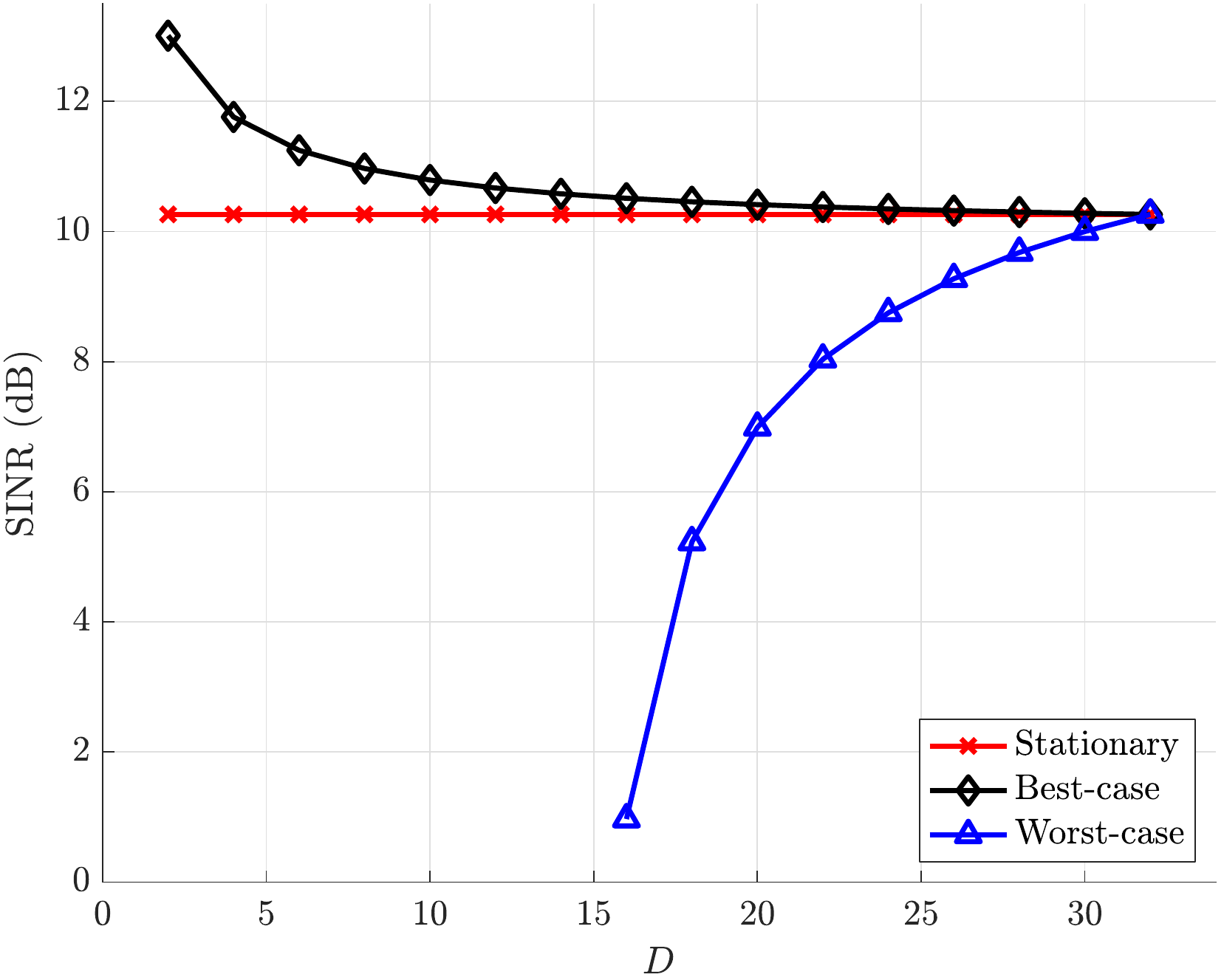}
        \caption{The SINR of $k$th user~vs the active number of antennas $D$ ($M=256,~K=64$, and $\rho=\SI{10}{\decibel}$).}
        \label{fig:ResAna}
\end{figure}

In Fig.~\ref{fig:ResAna}, we provide an example result to demonstrate the impact of non-stationarity on the performance of ZF precoding. We plot the SINR results against the active number of antennas per terminal, i.e., $D$. The number of users is 64. The array is linear with 256 antennas spaced by half the wavelength. At 2.4 GHz, the dimension of the array is 16 meters. We can see that depending on the configuration (i.e., best-case or worst-case) the SINR~can be significantly higher/lower than the SINR~of the stationary channels. As expected, the differences are larger for smaller values of $D$. 

The non-stationarity captured using VRs and subsequent analysis shows that non-stationarity has a significant impact on the performance of a massive MIMO system. As such, it is imperative to understand this impact and to exploit it in designing massive MIMO systems.

\subsection{Hybrid beamforming}

Hardware and cost constraints make it challenging to connect all the antennas in a massive MIMO system with dedicated RF-chains and high-resolution ADCs. Therefore hybrid analog-digital architectures, where a few RF-chains are connected to a large number of antennas are suitable for massive MIMO systems. The hybrid analog-digital architectures keep the cost and complexity under control by using fewer RF-chains compared to the number of antennas but allow multi-terminal multi-stream precoding that is not possible using analog-only architectures. 

There are several possibilities for implementing hybrid analog-digital architectures. The more flexible (but complex to implement) architecture is fully-connected architecture, where all the RF-chains are connected to all the antennas. A simpler (but less flexible) architecture is partially connected architecture in which every antenna is connected to a subset of RF-chains. Recently, dynamic hybrid architectures are also considered that adapt to the channel, hence providing flexible yet simpler implementation~\cite{Park2017Dynamic}.

The dynamic hybrid analog-digital architectures can be particularly beneficial in non-stationary channels. Motivated by the VR-based channel model discussed in the last section, it can be argued that a simple dynamic hybrid analog-digital architecture is one in which only the antennas corresponding to the VR are connected to the RF-chains. This is feasible as the antennas outside the VR do not have significant channel power. Thus a low complexity dynamic architecture can potentially provide performance close to the fully digital system but at low hardware cost.

\begin{figure}[h!]
\centering
       \includegraphics[width=0.45\textwidth]{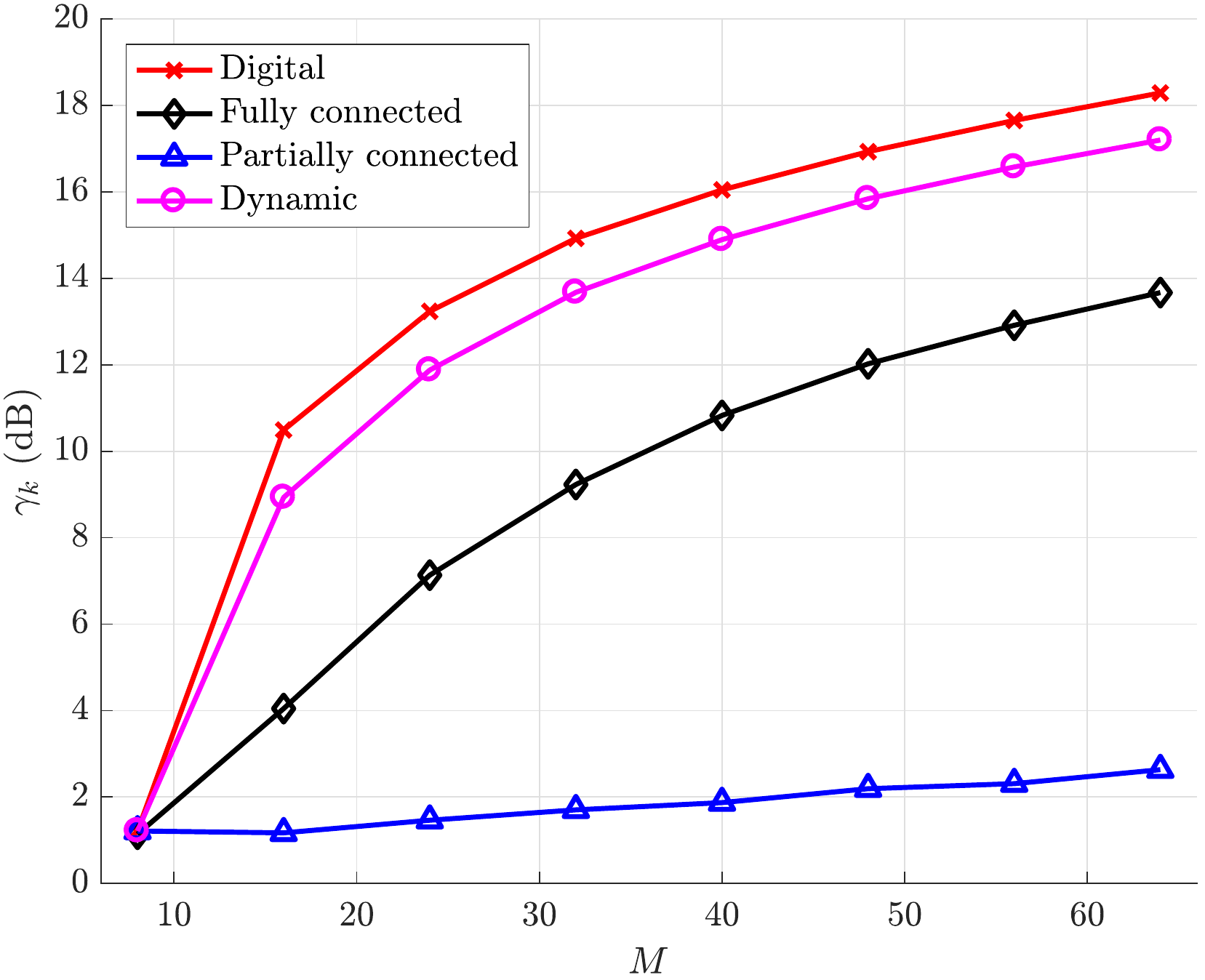}
        \caption{The SINR~vs the number of antennas $M$ ($D=M/2$, $K=16$, and $16$ RF-chains).}
        \label{fig:ResAna2}
\end{figure}

To show the benefit of non-stationarity aware system design, we provide simulation results. The array is linear with 256 antennas spaced by half the wavelength. At 2.4 GHz, the dimension of the array is 16 meters. Assuming $D=M/2$ size VR for each terminal (where visible antennas are chosen uniformly at random), we provide the average SINR of ZF precoder with different hardware architectures. There are $K=8$ terminals in the system. The fully digital architecture has $M$ RF-chains, and we use the algorithm proposed in~\cite{Liang2014Low} to obtain the hybrid precoders. The hybrid analog-digital architectures have RF-chains equal to the number of terminals $K$. The partially connected hybrid architecture has an RF-chain connected to $M/K$ successive antennas. The dynamic architecture has $k$th RF-chain connected only to the antennas visible to the terminal $K$. From the results in Fig.~\ref{fig:ResAna2} we can see that the dynamic architecture can provide performance better than fully-connected architecture (in non-stationary channels) and close to fully digital system.

Dynamic hybrid architectures are interesting for non-stationary massive MIMO. These architectures are an example of a system design that exploits the non-stationary nature of the massive MIMO channel. The results presented herein, however, are preliminary and a lot of research is required for practical designs. One major challenge in dynamic hybrid architectures is the efficient acquisition of channel state information (the presented results are based on genie aided CSI).

\subsection{Low complexity transceivers} \label{sec:lowComplexity}

One obvious consequence of having an extremely large number of antennas at the base station is its high complexity architecture. Even with simple linear transceivers, the base station should perform a large number of complex operations. This problem gets even worse when it comes to crowded scenarios with many terminals in the system. Therefore, implementing low complexity techniques is one major challenge. One possible way is to adapt the transceiver design to the non-stationary energy patterns of the terminals, complemented by distributed processing methods such as sub-array based architectures.
To determine low complexity transceivers, acquiring information about the VRs is critical. 

The existence of  VRs is the basis to implement low complexity linear transceivers such as the ZF. Indeed, the computational cost of implementing a ZF operation is dominated by the inversion of a matrix that has a band structure due to the VRs and might even be sparse.  However, implementing distributed techniques is more favorable due to lower complexity and more flexibility.  
Distributed processing is motivated not only by the computational cost but also by the ease of installation of very large arrays that are made out of smaller sub-arrays. Each sub-array carries out local processing of the signals while a central unit is responsible for the final data fusion step. 

As the terminals are connected to a subset of sub-arrays, a graph can be used to describe the connections between terminals and sub-arrays. When the terminals are connected to a small number of sub-arrays, the graph is sparse and becomes a  convenient tool to facilitate low complexity transceiver designs. 
Compared to a fixed sub-array division, a dynamic division leads to a better performance outcome where the division fits ideally the multi-terminal VR patterns and should be updated for the changes in the VR patterns. Simple learning algorithms can help in tracking the power pattern of the terminals over the array.
 
Considering the uplink, a linear fusion of the sub-array output signals is carried out at the central unit. When arrays are deployed in a very large structure, such as around the roof of a stadium, the processing can be structured hierarchically with a multi-stage fusion involving a hierarchical subset of sub-arrays at each step. 

Non-linear processing can be beneficial in some situations to improve performance compared to linear fusion. Due to spatial non-stationarities, multi-terminal interference patterns vary over the array so that one terminal experiences different interference conditions at each of the sub-array. Therefore, it becomes beneficial to detect a terminal from the sub-array with favorable interference conditions and then remove its contributions from the other sub-arrays, enhancing the signal to interference ratio of all the other terminals. This nonlinear method follows the principle of successive interference cancellation technique and was tested in~\cite{amiri2018extremely}.
More advanced receiver based on message passing among sub-arrays can be employed to reduce the performance gap with the optimal methods such as maximum likelihood.  

\begin{figure} 
\centerline{
\includegraphics[angle=270,origin=c,width=1\linewidth,trim={6cm 3.5cm 5cm 3cm },clip]{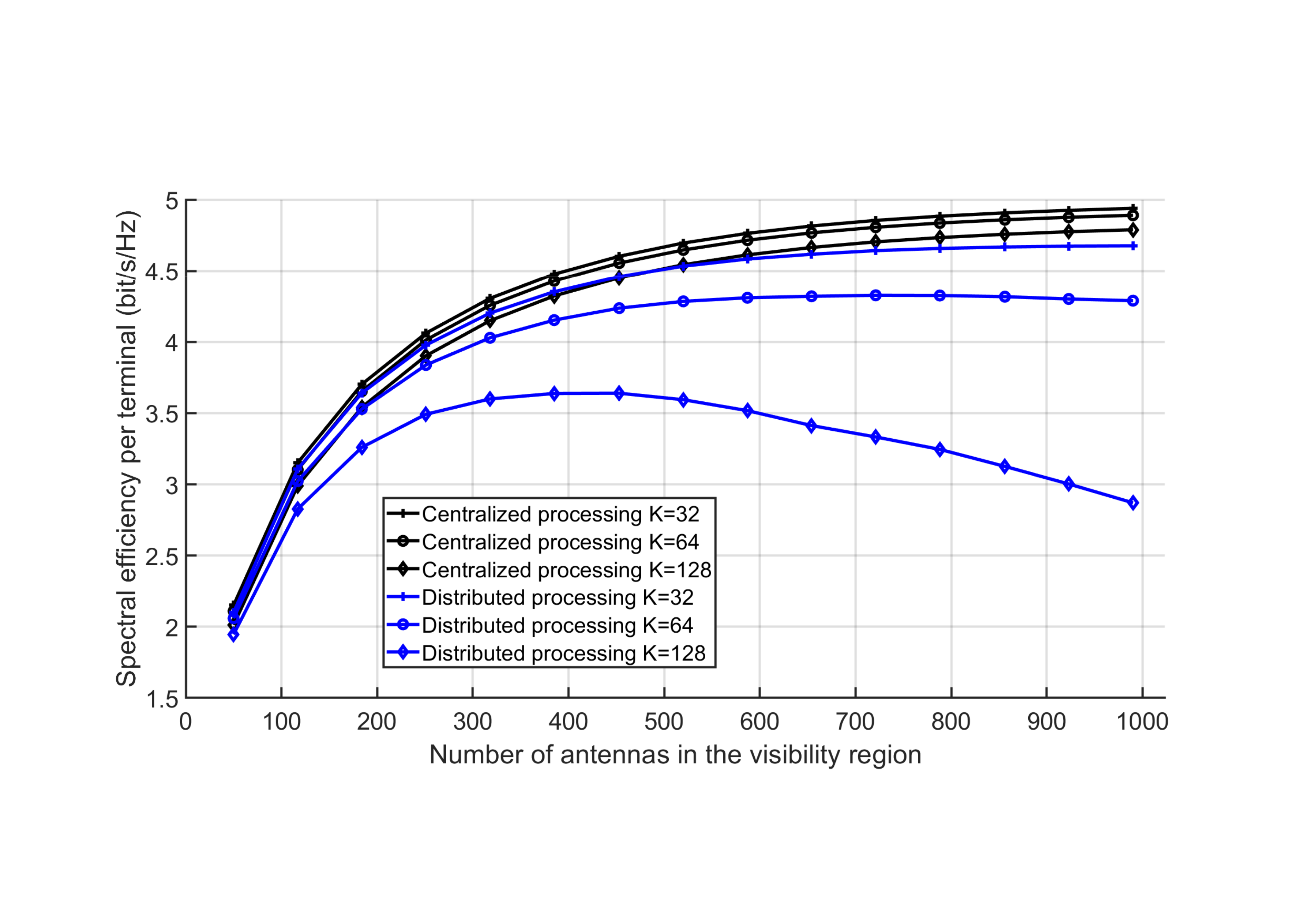}}\vspace{-2.2cm}
\caption{Rate per user comparison between centralized and distributed ZF processing vs number of contributing antennas in the VR of each terminal. ($M=1024$ and SNR= $15$dB).}
\label{fig:comparing_cent_vs_dist}
\end{figure}

In Fig.~\ref{fig:comparing_cent_vs_dist}, we test the notion of VR in a multi-terminal processing. The terminals are uniformly distributed in front of a linear array comprising 1024 antennas. At 2.4 GHz, the dimension of the array with antennas spaced by half the wavelength is 64 meters. The channel is assumed to have a Gaussian distribution while the energy variations along the array come from the path loss. The figure displays the spectral efficiency per terminal as a decreased number of antennas is considered in the VR of each terminal. 
For this channel model, we observe a saturation in the performance of the centralized processing. With some performance degradation, the processing can be reduced to a relatively small number of antennas per terminals (two sub-arrays of 128 antennas).
In distributed processing, we have 128 antennas per sub-array. We observe a degradation when the number of antennas in the VR  increases for a loaded system (128 terminals). As the VR becomes larger, the number of terminals to be processed per sub-array increases until reaching a regime where the sub-optimality of distributed processing becomes apparent. 

\section{Next steps} \label{sec:next}

\subsection{Characterizing the channel} \label{sec:characterizing}

Near-field channel measurements involving extremely large arrays where non-stationary patterns are visible are scarce (see section~\ref{fig:models}). Yet, they are necessary, as little is known about their non-stationary features outside of a theoretical framework. Channel measurements are needed to understand in more depth the propagation behavior in real-life set-ups. For example, it appears important to uncover how the wireless channel behaves in a large indoor venue with very large antenna panels deployed along the walls: is the channel sparse or does it demonstrate rich scattering?
The issue is not only about the phase front that is spherical and not planar anymore. It is also about the channel energy variations along the array as was the focus in this article. Those variations are the results of the path loss in line-of-sight but also the geometry of the building and reflecting structures (ceiling, floor, stairs, various objects) that are near the communicating panels and the end-terminals. 

{To characterize the channel, a multitude of measurements are needed in different deployment scenarios (e.g., different room types, outdoor scenarios) to guarantee statistical significance. Many more measurements are necessary to extract the non-stationary channel attributes but also other important features impacting channel modeling.
 An extremely large array views the propagation environment with super-resolution. The objects along the propagation do not look the same when illuminated by a large array. For example, large arrays can differentiate a set of reflecting entities that would be part of a cluster otherwise. Hence, the definition of clusters can be questioned, as well as the distribution of the small scale fading along the propagation path.
 Another example is about the modeling of large scale fading. 
As a terminal moves locally, large scale fading remains identical for a compact array size. With very large arrays, even very small movements might impact large scale quantities.}

\subsection{Embracing electromagnetics} \label{sec:EM}

Communication in the near-field and hence spherical wave modeling implies that the propagation features are dependent on the relative position of the end-terminals to the electromagnetic panels, the size of the panels, as well as their magnetic properties. This paper has considered very simplified models to highlight the impact of energy variations along the array panels. There is a need, though, to revisit communication theory to incorporate those electromagnetic attributes more faithfully. 

The incorporation of advanced electromagnetic features impacts the development of algorithms and likely adds to their complexity. As an example, compressed sensing methods are employed widely for sparse channel estimation. Those methods rely usually on a dictionary, i.e., typically an over-complete set of vectors that span the propagation space. Spherical waves imply that more parameters are necessary to describe the dictionary. 

Compared to stationary massive MIMO, the array aperture offers an additional degree of freedom: the assignment of a subset of antennas to each terminal. In section \ref{sec:lowComplexity}, 
we have restricted the processing area per terminal to the visibility region. Computational cost motivates the shrinkage of the processing area. Inside the processing area, the signal of a terminal is a signal of interest while it is treated as interference outside. Satisfying specific metrics provides another motivation: a terminal requiring more data is assigned a larger area while a fairness criterion might lead to a balanced assignment. This processing leads to a system model that lies between full network MIMO (the processing areas correspond to the visibility regions) and MIMO interference channel (the processing areas to all terminals are disjoint). This is reminiscent of the access point association in distributed settings. It is different though due to the high resolution in the assignment problem and an assignment that could be highly dynamic and follows the movements of the terminals. 
\section{Conclusions}
{XL-MIMO is an extreme but practical case of massive MIMO with larger apertures. This paper has focused on discrete antenna arrays of extremely large dimension that are deployed as part of a new large building structure. Along with active and passive large electromagnetic surfaces, they participate in a vision of ubiquitous connectivity where a connection is not achieved through access points anymore but rather through diffuse access that is located much closer to the end terminals. Such a vision is not realized yet and comprises many practical challenges. We have discussed how non-stationarities along the array found in XL-MIMO change the performance of MIMO systems and how visibility regions can be accounted for to decrease the computational load associated to MIMO transceivers in centralized or distributed implementations. When communication happens in the near-field, many other communication aspects are impacted.  For example, propagation attributes become different from the conventional far-field so that channel models need to be properly adjusted calling for new measurements. Directional beamforming is more complex because the beam does not depend on the directions only but also on the position of the terminals relative to the array. Those might be well-known properties of near-field communications. However, the array dimension brings specific challenges in terms of computational load that need to be addressed. }
\bibliographystyle{IEEEtran}
\bibliography{HeathMmWaveRefs}
\end{document}